\documentclass[conference]{IEEEtran}

\usepackage{cite}
\usepackage{amsmath,amssymb,amsfonts}
\usepackage{algorithmic}
\usepackage{graphicx}
\usepackage{textcomp}
\usepackage{xcolor}
\usepackage{hyperref}
\usepackage{authblk}
\usepackage{arydshln}
\usepackage{comment}

\begin{document}

\title{Loss functions incorporating auditory spatial perception in deep learning -- a review
}


\author[1]{Boaz Rafaely}
\author[2]{Stefan Weinzierl}
\author[1]{Or Berebi}
\author[2]{Fabian Brinkmann}
\affil[1]{School of Electrical and Computer Engineering, Ben-Gurion University of the Negev, Beer-Sheva, Israel}
\affil[2]{Audio Communication Group, Technische Universität Berlin, Berlin, Germany}

\maketitle

\begin{abstract}
Binaural reproduction aims to deliver immersive spatial audio with high perceptual realism over headphones. Loss functions play a central role in optimizing and evaluating algorithms that generate binaural signals. However, traditional signal-related difference measures often fail to capture the perceptual properties that are essential to spatial audio quality. This review paper surveys recent loss functions that incorporate spatial perception cues relevant to binaural reproduction. It focuses on losses applied to binaural signals, which are often derived from microphone recordings or Ambisonics signals, while excluding those based on room impulse responses. Guided by the Spatial Audio Quality Inventory (SAQI), the review emphasizes perceptual dimensions related to source localization and room response, while excluding general spectral–temporal attributes. The literature survey reveals a strong focus on localization cues, such as interaural time and level differences (ITDs, ILDs), while reverberation and other room acoustic attributes remain less explored in loss function design. Recent works that estimate room acoustic parameters and develop embeddings that capture room characteristics indicate their potential for future integration into neural network training. The paper concludes by highlighting future research directions toward more perceptually grounded loss functions that better capture the listener's spatial experience.
\end{abstract}


\section{Introduction}

Binaural synthesis creates immersive audio experiences through headphones by simulating how sounds reach the ears in real acoustic environments. These systems rely on accurately rendering spatial auditory cues, often using signal processing methods that incorporate head-related transfer functions (HRTFs). As binaural audio becomes increasingly important in applications such as virtual reality, gaming, hearing aids, and telepresence, and as deep learning methods become more widely adopted in audio processing, there is a growing need for suitable loss functions—metrics used to guide the training and evaluation of learning-based algorithms. In the context of binaural audio, effective loss functions must go beyond conventional signal accuracy and incorporate perceptually relevant aspects of spatial hearing, ensuring that the optimized outputs preserve realistic and spatially accurate auditory experiences.

Recent advances in spatial audio research have increasingly emphasized perceptual aspects, particularly through structured descriptors such those summarized in the Spatial Audio Quality Inventory (SAQI) \cite{lindau2014SAQI}. This reflects a broader trend toward integrating psychoacoustic insights into the design and processing of spatial audio systems \cite{brandenburg2019perceptual,Hacihabiboglu2017Review}. Several recent studies have examined perceptually relevant attributes of spatial hearing, such as localization, envelopment, externalization, and reverberation perception, and have explored how these attributes can be evaluated using auditory models \cite{Llado2025}. Other recent efforts have reviewed the use of machine learning for HRTF individualization \cite{Fantini2025} and proposed data-driven pipelines for spatial audio capture and reproduction \cite{cobos2022Review}. Nevertheless, as emphasized in earlier work on spatial audio perception \cite{BRADLEY2011,Weinzierl2018RAQI}, perceptual qualities related to reverberation and room acoustics play a critical role in spatial realism. Despite this, such attributes remain underrepresented in the design of current loss functions for learning-based spatial audio processing.

\begin{figure}[h]
    \centering
    \includegraphics[width=0.99\linewidth]{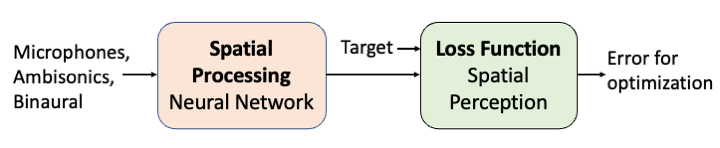}
    \caption{The framework representing the signal processing chain and the incorporation of loss functions. Examples of signal processing tasks and methods within this framework are presented in the recent review paper by Rafaely et al.~\cite{rafaely2022Review}, whereas the loss functions are discussed in the present paper.}
    \label{fig:framework}
\end{figure}

This paper reviews loss functions that explicitly incorporate spatial perception attributes and evaluates their relevance to binaural reproduction. The focus is on loss functions applied to binaural signals, or signals derived from microphone arrays or Ambisonics representations, as illustrated in Fig.~\ref{fig:framework}. We exclude metrics designed solely for characterizing systems, such as room impulse responses of transfer functions. Guided by the SAQI framework, the review identifies key limitations in current approaches—particularly the limited treatment of reverberation and room acoustics—and outlines opportunities for developing loss functions that are more closely aligned with perceptual spatial audio cues.

\section{Spatial audio qualities}

The Spatial Audio Quality Inventory (SAQI,~\cite{lindau2014SAQI}) contains perceptual qualities relevant to the evaluation of audio in virtual and augmented reality. It can serve as a starting point for identifying qualities for which loss functions are desirable. The qualities in SAQI are divided into eight categories. For the scope of this article, the current review focuses on selected qualities from two categories: \emph{geometry}, containing qualities related to the spatial perception of the sound source; and \emph{room}, containing qualities related to the spatial perception of the acoustic environment. A large set of room acoustic qualities was further developed in the Room Acoustic Quality Inventory (RAQI,~\cite{Weinzierl2018RAQI}). While a subset of these qualities is also contained in SAQI, other high-level qualities, such as the `ease of listening' and `liveliness', are excluded from the current work.

In addition to the above, we included the (overall) difference as an integrative quality and summarize the eight SAQI qualities given in the \emph{timbre} category as a single quality, termed `coloration'. Strictly spoken, these are not spatial qualities, but we included them because they are important qualities for evaluating spatial audio in general. Coloration, which we consider to be mostly an attribute of the sound source, has an outstanding importance to the overall quality of (spatial) audio in general~\cite{Rumsey2005,LeBagousse2012}. Table~\ref{tab:Qualities} lists the selected qualities, and the reader is referred to Lladó et~al.~\cite{Llado2025} for a more comprehensive review of the SAQI qualities, related measures, and available auditory models.

The overview in Table~\ref{tab:Qualities} shows that a relatively small number of technical measures is connected to the selected spatial audio qualities. While these measures could (mostly) be implemented as loss functions, their exact relation to the corresponding perceptual quality is often non-trivial. Many measures associated with source-related qualities are used as building blocks of more elaborated auditory models, for example for sound source localization~\cite{Majdak2022}; however, these may not be implementable as loss functions. In other cases, the Just Noticeable Difference (JND) of the measures is not fully understood~\cite{BRADLEY2011}, which makes applying them in a perceptually meaningful manner difficult. These factors make developing loss functions tightly linked to perceptual qualities challenging.

\begin{table}[h]
    \centering
    \begin{tabular}{|c|p{1.5cm}|p{3cm}|p{1.75cm}|}
    \hline
    \textbf{Object}&\textbf{Quality}  & \textbf{Description} & \textbf{Measures} \\
    \hline
    $\mathcal{S,R}$&Difference &Existence of a noticeable difference.& All of the below\\
    \hline\hline
    $\mathcal{S}$&Coloration & Loudness differences between a reference and test stimulus, integrated across ears and frequency & (Differences of) spectral cues\\
    \hline\hline
    $\mathcal{S}$ &Horizontal direction & Direction of a sound source in the horizontal plane & ITD, ILD \\
    \hdashline[1pt/1pt]
    $\mathcal{S}$ &Vertical direction & Direction of a sound source in the vertical plane &  Spectral cues \\
    \hdashline[1pt/1pt]
    $\mathcal{S}$ &Front-back position & Position of a source in front or behind the listener & Spectral cues  \\
    \hdashline[1pt/1pt]
    $\mathcal{S}$ &Width, height, depth  &  Perceived extent of a source in the horizontal, vertical and radial directions  &  ASW, IACC \\
   \hdashline[1pt/1pt]
    $\mathcal{S}$ &Distance & Perceived distance of a source  & Loudness, DRR     \\
   \hdashline[1pt/1pt]
    $\mathcal{S}$ &Externali- zation & Perception of a source within or outside the listener's head & ILD (fluctuation), spectral cues  \\
   \hdashline[1pt/1pt]
    \hline\hline
    $\mathcal{R}$ &Level &   Strength of reverberant sound component & G, T60, EDT, DRR \\
    \hdashline[1pt/1pt]
    $\mathcal{R}$&Duration  &  Duration of the reverberant decay. Well audible at the end of signals & T60, EDT, DRR \\
    \hline
    \end{tabular}
    \vspace{1mm}
    \caption{Spatial audio qualities with their description and measures sorted by the objects to which the pertain ($\mathcal{S}$: Source and $\mathcal{R}$: Room). Abbreviations: Interaural time and level differences (ITD, ILD), apparent source width (ASW), interaural cross-correlation (IACC), direct-to-reverberatn ratio (DRR), sound strength (G), reverberation time (T60), early decay time (EDT), listener envelopment (LEV), and lateral fraction (LF).}
    \label{tab:Qualities}
\end{table}

\section{Loss functions in spatial audio processing with neural networks}

Designing effective loss functions is central to training deep neural networks in general and for spatial audio tasks in particular. A loss function quantifies the discrepancy between the model's output and the desired target. Its choice critically affects convergence, generalization, and in particular here, the perceptual relevance of the learned model. For use in gradient-based optimization, a loss function must be differentiable almost everywhere and expressed in terms of elementary operations whose derivatives are well-defined. This requirement ensures compatibility with automatic differentiation frameworks, such as PyTorch or TensorFlow, which rely on the chain rule to compute gradients efficiently~\cite{paszke2019pytorch}. It is worth noting that some commonly used operations, such as the absolute value and Rectified Linear Unit (ReLU) functions, are non-differentiable at zero. Nevertheless, automatic differentiation engines assign subgradient values (e.g., 0 or 1 for ReLU) to enable training. This approach can introduce optimization artifacts, but it is generally effective in practice~\cite{pytorch2024}.

With this in mind, the most relevant publications for this review paper are those that describe a loss function incorporating spatial perception qualities that have actually been used for neural network training. Nevertheless, publications may include perceptually motivated measures and estimation of parameters that may be important for perception, but that have not been employed as training loss. Accordingly, we present the following classification of loss functions according to their applicability as spatial perception-based losses in the context of neural network training.

\begin{itemize}
    \item \textbf{SL-NN (Spatial Loss for Neural Networks):} 
    Loss functions that explicitly incorporate spatial perception qualities and have been used as training objectives for neural networks.

    \item \textbf{EST-NN (Estimators with Neural Networks):} 
    Neural networks trained to estimate spatial audio measures (e.g., T60, Distance), though not directly used as loss functions. These networks could potentially be adapted for use in training due to their learned structure.

    \item \textbf{EMB-NN (Embeddings for Neural Networks):} 
    Representations learned by neural networks that encode audio signals in a way that may implicitly capture spatial qualities. While not estimating explicit spatial parameters, such embeddings can be used to assess the similarity of signals with respect to spatial characteristics, offering potential for use in loss design.

    \item \textbf{SL-CAN (Spatial Loss Candidate):} 
    Loss functions or perceptually motivated measures that incorporate spatial perception criteria but have not yet been used for neural network training. While not currently employed as loss terms, they show potential for such use based on their relevance to spatial audio perception. 

\end{itemize}

\noindent \\In the following, we review loss functions that are available for the spatial audio qualities listed in Table~\ref{tab:Qualities} and summarize the findings in Tables~\ref{tab:papers-src} and~\ref{tab:papers-room}.

\begin{table}[h]
    \centering
    \begin{tabular}{|p{12mm}|p{15mm}|p{13mm}|p{11mm}|p{15mm}|}
    \hline
    \textbf{Ref}   &  \textbf{Name} & \textbf{CH} & \textbf{Class} & \textbf{Quality} \\
    \hline
    \cite{Jiang2023a} & GML & $2$ & SL-CAN & Difference \\
    \hdashline[1pt/1pt]
    \cite{manocha2022saqam, Zheng2025} & SAQAM, HAPG-SAQAM & $2$ & SL-CAN & Difference, Coloration, Spatial Quality, Localization \\
    \hdashline[1pt/1pt]
    \cite{eurich2024computationally} & eMoBi-Q & $1,\, 2$ & SL-CAN & Difference, Localization \\
    \hdashline[1pt/1pt]
    \cite{Schafer2013,Delgado2023}& binaural PEAQ & $2$ & SL-CAN& Difference\\
    \hdashline[1pt/1pt]
    \cite{flessner2017assessment,Flessner2019} & BAM-Q & $2$ & SL-CAN & Difference, Localization\\
    \hdashline[1pt/1pt]
    \cite{Narbutt2018,Narbutt2020} & AMBIQUAL & FOA, HOA & SL-CAN & Difference, Localization\\
    \hdashline[1pt/1pt]
    \cite{Fantini2025,Hogg2025} & LSD & $2$ & SL-NN & Coloration \\
    \hdashline[1pt/1pt]
    \cite{Yao2024a} & & $2$ & SL-CAN & Coloration\\
    \hdashline[1pt/1pt]
    \cite{McKenzie2025} & PBC-2 & $2$ & SL-CAN & Coloration \\
    \hdashline[1pt/1pt] 
    \cite{McKenzie2025} & BSD$_\mathrm{A}$ & $2$ & SL-CAN & Coloration \\
    \hdashline[1pt/1pt]
    \cite{Krause2021} & & $2$ & EST-NN & Localization \\
    \hdashline[1pt/1pt]   
    \cite{panah2025binaqual} & BINAQUAL & $2$ & SL-CAN & Localization\\
    \hdashline[1pt/1pt]
    \cite{manocha2021dplm} & DPLM & $2$ & EST-NN & Localization\\
    \hdashline[1pt/1pt]
    \cite{francl2022deep} &   & $2$ & EST-NN & Localization\\
    \hdashline[1pt/1pt]
    \cite{berebi2023imagls} & iMagLS & FOA & SL-CAN & Localization \\
    \hdashline[1pt/1pt]
    \cite{berebi2024feasibility} & BSM-iMagLS & $4,\,5,\,6,\,12$ & SL-CAN & Localization \\
    \hdashline[1pt/1pt]
    \cite{berebi2025bsm} & BSM-iMagLS & $4,\,5,\,6,\,12$ & SL-NN & Localization \\
    \hdashline[1pt/1pt]
    \cite{10889034} & MMagLS & FOA & SL-NN & Localization \\
    \hdashline[1pt/1pt]
    \cite{tokala2024binaural} &  & $2$ & SL-NN & Localization \\
    \hdashline[1pt/1pt]
    \cite{neudekinvestigation} & & $2$ & EST-NN & Distance \\
    \hdashline[1pt/1pt]
    \cite{gburrek2024diminishing} & & $2$ & EST-NN & Distance \\
    \hdashline[1pt/1pt]
    \cite{neri2024speaker} & & $1$ & EST-NN & Distance \\
    \hline
    \end{tabular}
    \vspace{1mm}
    \caption{Review of papers incorporating loss functions related to perception of sound sources.}
    \label{tab:papers-src}
\end{table}

\subsection{Loss Functions Addressing Overall Perceptual Differences}\label{sec:overall_difference}

This subsection reviews loss functions and related models that aim to capture overall perceptual differences between reference and test signals, without targeting specific spatial attributes. These approaches typically rely on learning-based predictors or perceptually motivated metrics to estimate human judgments of audio quality similar to MUSHRA (MUltiple Stimuli with Hidden Reference and Anchor~\cite{ITU-R1534-3}) scores. Note that the terms `quality' and `difference' are often used interchangeably. A difference of zero is equivalent to the maximum achievable quality, and a quality of zero is equivalent to the maximum possible difference. While originally developed for codec evaluation or general audio degradation assessment, several of these methods have shown potential for broader application in binaural and spatial audio contexts.

The Generative Machine Listener (GML,~\cite{Jiang2023a}) is a deep neural network that was trained to predict MUSHRA scores based on Gammatone spectrograms of stereo or binaural test and reference signals. Trained on a large dataset of listening tests involving codec-degraded audio, it achieved a Spearman rank correlation of 0.91 with mean scores. Although the GML is proprietary, its output could potentially serve as a perceptually informed loss function for optimizing spatial audio algorithms putting it into the SL-CAN category.

The Spatial Audio Quality Assessment Metric (SAQAM~\cite{manocha2022saqam}) is a deep neural network that estimates listening, spatial, and overall quality based on reference and test binaural signals of flexible lengths and audio content. Quality estimates are based on the distances between feature embeddings, which are computed using an embedding neural network (EMB-NN). SAQAM was trained without perceptual ratings, using speech signals that were spatialized with a large set of binaural room impulse responses and degraded through noise, distortion, and compression. Despite this, the model showed a moderate-to-strong correlation with human ratings from five listening tests (e.g., 0.79 for overall quality). Although not originally designed as a loss function, SAQAM’s output could potentially guide perceptually informed optimization in spatial audio systems, making it an SL-CAN.

The Human Auditory Perception Guided SAQAM (HAPG-SAQAM~\cite{Zheng2025}) extends the original SAQAM by incorporating additional perceptually relevant features, including gammatone frequency cepstral coefficients, while discarding less relevant information through an attention mechanism based on cross-channel comparisons. It also applies optimized weighting of listening and spatial quality derived from listening test data to produce an overall quality estimate. These enhancements improved the model’s correlation with perceptual ratings, reaching 0.83 for overall quality and 0.77 for spatial quality. Given its strong alignment with perceptual judgments, HAPG-SAQAM is a promising candidate for use as a loss function that reflects overall and spatial audio quality, thus belonging to the SL-CAN category.

These models were primarily developed for evaluating the quality of lossy audio codecs. In the case of SAQAM and HAPG-SAQAM, the models were trained exclusively on speech signals at a sampling rate of 16\,kHz. Nonetheless, the results suggest they can generalize across audio content, degradation types, and sampling rates. To the best of our knowledge, none of these models are publicly available. In their absence, similar approaches could be re-engineered, or non-differentiable algorithms (classified as SL-CAN) could be adapted for use as loss functions (SL-NN). Examples include eMoBi-Q~\cite{eurich2024computationally}, binaural extensions of PEAQ~\cite{Schafer2013,Delgado2023}, BAMQ~\cite{flessner2017assessment,Flessner2019}, and Ambisonics Quality (Ambiqual)~\cite{Narbutt2018,Narbutt2020}, which are at least partially available for research use.

\subsection{Loss Functions Preserving Sound Color}

Coloration is typically estimated by measuring spectral differences between a processed and a reference signal. In most applications involving HRTF processing, such as individualization~\cite{Fantini2025} or spatial upsampling~\cite{Hogg2025}, the Log Spectral Distortion (LSD) measure is used. LSD calculates the root mean square error in decibels across frequency bins, often averaged across source positions and ears. While its simplicity makes it a convenient loss function of the \mbox{SL-NN} category, LSD has been shown to correlate poorly with perceived coloration, reaching Pearson correlations of $\rho=0.59$~\cite{Yao2024a} and $\rho=0.67$~\cite{McKenzie2025}. This may be due to its neglect of the non-linear frequency resolution of human hearing, and the limited perceptual relevance of bin-wise spectral differences, particularly in reverberant binaural signals, where comb-filter artifacts may be inaudible.

Improved predictors of coloration account for the nonlinear frequency resolution of the human auditory system by computing spectral differences in auditory bands with perceptual smoothing. Notable examples include the Perceptually Enhanced Spectral Distance Metric, $\rho=0.79$~\cite{Yao2024a}, the Auditory Basic Spectral Difference (BSD$\mathrm{A}$, $\rho=0.84$~\cite{McKenzie2025}), and Predicted Binaural Colouration 2 (PBC-2, $\rho=0.92$~\cite{McKenzie2025}), all of which fall into the SL-CAN category. Note that the correlations above should be compared with care because different audio stimuli were used in the corresponding studies. Among these measures, BSD$_\mathrm{A}$ can be implemented as a loss function, similar in form to the iMagLS loss~\cite{berebi2023imagls}. Alternatively, overall spatial audio quality metrics, such as those discussed in Section~\ref{sec:overall_difference}, may also serve this purpose given the strong influence of coloration on perceived audio quality~\cite{Rumsey2005,LeBagousse2012}.

\subsection{Loss Functions Addressing Perceived Source Localization}

Accurate modeling of sound source localization is a key objective in spatial audio. Human directional hearing relies primarily on ITD and ILD for localization in horizontal direction, while spectral cues play a central role in vertical localization and front–back discrimination. Recent work has explored how these perceptual cues can be integrated into neural networks, evaluation metrics, and optimization pipelines to improve spatial accuracy in binaural reproduction.

Several recent studies have incorporated localization cues directly into the loss functions of neural networks, thereby making spatial accuracy an explicit training objective. For instance, HAPG-SAQAM~\cite{Zheng2025} employs a neural network trained on binaural signals degraded in various ways to predict spatial audio quality. The model is optimized using a composite loss that combines a triplet embedding objective with a perceptual weighting function derived from listener ratings. This loss promotes spatially meaningful embeddings by emphasizing differences in ILD and ITD between signals, making the approach an example of an SL-NN class.

Ambisonics Binaural Rendering via Masked Magnitude Least Squares (MMagLS)~\cite{10889034} targets vertical localization by emphasizing spectral notches associated with pinna cues. The model processes first-order Ambisonics and is trained to generate binaural outputs that preserve elevation cues through a frequency-dependent spectral mask. This mask highlights regions containing pinna-related notches, resulting in an enhanced modeled localization in the median plane. As a neural network trained with a perceptually motivated loss function, MMagLS is classified as an SL-NN.

Tokala et al.~\cite{tokala2024binaural} proposed a deep complex convolutional transformer network for binaural speech enhancement, using two-channel STFT inputs. Their model is based on a hybrid convolutional encoder-decoder and transformer architecture and is trained with a four-part loss combining Signal-to-Noise Ratio (SNR) and Short-Time Objective Intelligibility (STOI) objectives with explicit ILD and interaural phase difference (IPD) preservation terms. By restricting interaural cue penalties to speech-active time–frequency regions, the loss directly enforces spatial accuracy in the enhanced output. This makes the approach a clear example of the SL-NN category.

In contrast to loss-based approaches, several works use ILD and ITD as input features to neural networks without incorporating these cues directly into the loss function. These models are trained to estimate perceptual or spatial attributes and are thus classified as estimators trained with neural networks (EST-NN). For example, DPLM~\cite{manocha2021dplm} is a convolutional network trained to predict perceptual localization error from time–frequency representations of ILD and ITD. The network is optimized using a regression loss to match subjective localization ratings, rather than enforcing spatial accuracy through the loss itself.
A similar approach is taken by Krause et al.~\cite{Krause2021}, who trained a network to classify the direction and proximity of overlapping sound events from binaural input. The model uses ILD and ITD features extracted from the recordings and is trained with a cross-entropy loss to predict categorical labels such as direction (left, right, front, back) and proximity (near or far). These examples demonstrate how interaural cues can inform network predictions without being explicitly integrated into the loss function.

A broader class of works defines perceptually motivated spatial loss functions that are not currently used in neural network training. These include full-reference objective metrics, signal-processing optimization criteria, and lightweight predictive models that rely on spatial cues. For example, BINAQUAL~\cite{panah2025binaqual} assesses localization similarity in binaural audio by comparing time–frequency representations of ILD, ITD, and interaural coherence between reference and test signals. While originally designed as a standalone evaluation metric, its structure can be adapted into a differentiable loss, placing it within the SL-CAN category.
Similarly, BAM-Q~\cite{flessner2017assessment} extracts ILD and ITD features from binaural signals and uses them in a regression model to predict perceived spatial quality. Although effective as an evaluation metric, it has not been applied in training, so it falls under the SL-CAN category.

A comprehensive study by Francl et al.~\cite{francl2022deep} explored sound localization from a behavioral perspective using deep neural networks trained on simulated binaural recordings. The networks received input in the form of cochleagram-like representations derived from stereo signals processed with HRTFs. The networks were trained as classifiers to predict discrete azimuth and elevation bins corresponding to the source location, thereby implicitly optimizing spatial accuracy through a cross-entropy loss. Though the loss function does not encode ILD or ITD directly, the networks demonstrated strong sensitivity to these cues, replicating classic human localization behaviors such as frequency-dependent ITD/ILD reliance, spectral cue usage for elevation, and robustness to spectral smoothing. Since spatial cues are not explicitly embedded in the loss function, this work is best categorized as EST-NN: the model estimates the
direction of the source using spatially informative input, but the training objective does not enforce perceptual localization criteria directly.

Beyond binaural similarity metrics, several signal processing approaches incorporate ILD directly into their optimization objectives. iMagLS~\cite{berebi2023imagls} extends the MagLS formulation~\cite{zotter2019ambisonics} for first-order Ambisonics HRTF optimization by adding a penalty term for ILD deviations, improving alignment of lateral localization cues in Ambisonic-to-binaural rendering. A related study by Berebi et al.~\cite{berebi2024feasibility} applies a similar strategy to filter design for binaural rendering from microphone array inputs, again minimizing ILD error as part of the objective. Although these methods do not involve neural networks, they exemplify perceptually motivated spatial loss design and fall within the SL-CAN category. More recently, BSM-iMagLS~\cite{berebi2025bsm} adopted a similar ILD-informed formulation to train a neural network for BSM using head-mounted microphone array inputs. This work is an example of how to implement the SL-CAN loss proposed in~\cite{berebi2024feasibility} within a neural network training pipeline by embedding the ILD error directly into the learning objective. As such, it represents an instance of the SL-NN category.

Finally, eMoBi-Q~\cite{eurich2024computationally} is a computationally efficient quality prediction model that incorporates ILD and interaural coherence within a compact feature set to estimate listener quality ratings. Operating on binaural input it uses regression to map features to perceptual scores. Although not designed as a training loss, its use of perceptually relevant spatial cues places it in the SL-CAN category.

Taken together, these works present a variety of strategies for integrating ILD, ITD, and spectral cues into spatial audio modeling. SL-NN approaches, such as HAPG-SAQAM and MMagLS, embed perceptual spatial attributes directly into the training objective. EST-NN methods use spatial cues as input features for predicting localization or directional attributes. Although SL-CAN approaches have not been used in training, they provide interpretable, perceptually grounded formulations that could be adapted into loss functions. These contributions underscore the increasing importance of perceptually informed modeling in spatial audio and point toward future loss functions that more closely align with the auditory mechanisms underlying sound localization.

\subsection{Loss Functions Addressing Perceived Source Distance}

Recent work on source distance estimation has used a range of loss function formulations, each reflecting different assumptions about the task. The choice of the loss function plays a key role in shaping the learning objective, the model’s behavior, and its ability to generalize. This section reviews representative approaches and highlights how loss functions have been used to preserve information about source distance.

Several studies have framed source distance estimation as a supervised regression task using standard loss functions such as mean squared error (MSE). Neudek et al.~\cite{neudekinvestigation}, for example, trained neural networks on simulated binaural signals using either raw waveforms or extracted features, with an MSE loss function calculated for the difference between predicted and true distances. However, the loss does not explicitly account for perceptual cues, such as direct-to-reverberant ratio (DRR) or intensity fall-off, which are expected to be learned implicitly. Thus, this approach is classified as EST-NN.

Gburrek et al.~\cite{gburrek2024diminishing} also approached distance estimation as a regression task, using mono audio and MSE loss. Their main contribution lies in the training data, generated using image source models combined with stochastic late reverberation to simulate near-realistic spatial scenes. Despite the improved data, the use of a standard, non-spatial loss places this work in the EST-NN category.

Neri et al.~\cite{neri2024speaker} proposed a composite loss function to capture the temporal dynamics of moving sources. A neural network was trained on multichannel audio to estimate speaker distance over time, using a combination of global MSE and a frame-wise regularization term that penalizes rapid changes in predictions. This approach promotes temporal consistency, which is perceptually relevant, however, the loss does not explicitly model distance cues. The approach remains in the EST-NN category as a partial step toward perceptually informed training.

An alternative formulation is offered by Krause et al.~\cite{Krause2021}, who framed the task as a classification problem. Their model uses binaural recordings as input to predict source direction and proximity (near or far) with discrete class labels. Training is performed using cross-entropy loss on proximity categories, which is a coarse yet robust approach to distance estimation. As it does not rely on explicit perceptual cues, this model is also classified as EST-NN.

In these works, source distance estimation is most often treated as a general-purpose regression or classification task, typically using standard losses such as MSE or cross-entropy. While effective for training, these losses do not explicitly model perceptual distance cues such as DRR, IACC, or intensity fall-off. Classification-based approaches offer robustness but limited resolution, and composite losses—such as those incorporating temporal smoothness—introduce structural elements heuristically rather than through perceptually grounded design. Examples for neural network-based estimators of room-related cues are given in the next section.  As a result, the reviewed approaches fall into the EST-NN category. This highlights a gap in the current literature and suggests an opportunity for future work to explore loss functions that explicitly reflect the auditory cues underlying distance perception. As spatial rendering and immersive audio systems advance, distance-aware training objectives will become increasingly important.

\subsection{Loss Functions Addressing Room Acoustics Perception}

Loss functions targeting room acoustics aim to capture perceptually relevant attributes such as reverberation level and decay (cf. Table~\ref{tab:Qualities}), which are central to spatial realism, yet have received limited attention in loss design. Most existing work instead focuses on using neural networks to estimate acoustic measures such as T60 and DRR, typically from single-channel speech, though some approaches also consider multi-channel input. While these models could potentially inform loss functions that reflect perceptual differences in room acoustics, their application in this context remains largely unexplored.

\begin{table}[h]
    \centering
    \begin{tabular}{|p{12mm}|p{9mm}|p{11mm}|p{12mm}|p{25mm}|}
    \hline
    \textbf{Ref} & \textbf{Name} & \textbf{CH} & \textbf{Class} & \textbf{Measure} \\
    \hline
    \cite{cox2001extracting,Gamper2018,Li2021,Li2023,Deng2020} & & 1 & EST-NN & T60 \\
    \hdashline[1pt/1pt]
    \cite{Genovese2019} & & 1 & EST-NN & Volume \\
    \hdashline[1pt/1pt]
    \cite{Mack2020} & & 1 & EST-NN & DRR \\
    \hdashline[1pt/1pt]  
    \cite{Eaton2016} & {ACE} & 1, 2, 3, 5, 8, 32 & EST-NN & T60, DRR \\
    \hdashline[1pt/1pt]
    \cite{Looney2020} & & 1 & EST-NN & T60, DRR \\
    \hdashline[1pt/1pt]
    \cite{Saini2023,saini2024end,Gotz2023a} & & 1 & EST-NN & T60, C50 \\
    \hdashline[1pt/1pt]
    \cite{Coldenhoff2024} & MOSRA & 5 & EST-NN & T60, DRR, C50 \\
    \hdashline[1pt/1pt]
    \cite{Meng2025} & & FOA & EST-NN & T60, DRR, C50 \\
    \hdashline[1pt/1pt]
    \cite{Srivastava2021} & & 2 & EST-NN & T60, Absorption, Volume, Area\\
    \hdashline[1pt/1pt]
    \cite{Wang2025} & BERP & 1 & EST-NN & T60, C50, C80, D50, EDT, Ts, Occupancy \\
    \hdashline[1pt/1pt]
    \cite{Bitterman2024} & RevRIR & 1 & EMB-NN & Room shape (small, large, hall) \\
    \hdashline[1pt/1pt]
    \cite{Gotz2023b} & & 1 & EMB-NN & T60, C50, Volume (small/large) \\
    \hdashline[1pt/1pt]
    \cite{Omran2023} & & 1 & EMB-NN & T60 \\
    \hdashline[1pt/1pt]
    \cite{Gotz2024} & & 1 & EMB-NN & T60, C50 \\
    \hdashline[1pt/1pt]
    \cite{dumpala24_interspeech} & XANE & 1 & EMB-NN & C50, T60, DRR, Volume \\
    \hline
    \end{tabular}
    \vspace{1mm}
    \caption{Papers incorporating neural networks and estimation of room acoustics attributes (EST-NN), or that include embeddings motivated by room acoustics (EMB-NN).}
    \label{tab:papers-room}
\end{table}

The first group of papers focuses on estimating a single room acoustic measure, typically the reverberation time T60, which is one of the most widely used descriptors of reverberant environments. These studies estimate T60 using single-channel reverberant speech. One of the earliest contributions in this area by Cox et al.~\cite{cox2001extracting} employed a fully connected neural network to estimate T60. Gamper and Tashev~\cite{Gamper2018} presented a convolutional neural network (CNN) that outperformed traditional algorithms from the ACE challenge~\cite{Eaton2016}, which will be discussed later. Subsequent work by Li et al.~\cite{Li2021} examined the use of both regression and classification approaches for T60 estimation and later applied this framework to dereverberation~\cite{Li2023}. Deng et al.~\cite{Deng2020} proposed a convolutional recurrent neural network (CRNN) capable of real-time estimation that generalizes across rooms and noise conditions. Together, these models illustrate the feasibility and accuracy of learning-based T60 estimation from reverberant speech. 

The next single-measure task addressed in~\cite{Genovese2019} involves estimating room volume from noisy, single-channel speech using a CNN-based regression model. The results suggest that reverberant speech carries sufficient information to support rough estimates of the physical room size. In another work, Mack et al.~\cite{Mack2020} focused on blind estimation of the direct-to-reverberant ratio (DRR), employing time-frequency masking to separate the direct and reverberant components. This method also uses a single-channel input and reflects another perceptually relevant acoustic attribute.

The ACE Challenge~\cite{Eaton2016} introduced blind estimation of both T60 and DRR, using single-channel and multi-channel microphone array recordings. These measures were estimated both as broadband scalars and in sub-bands, underscoring the importance of frequency-dependent analysis. While many algorithms participated, only a few utilized neural networks. Looney and Gaubitch~\cite{Looney2020} proposed a CNN-based model for jointly estimating T60 and DRR, along with signal-to-noise ratio, showing good generalization to ACE Challenge data.

Another group of papers addresses the joint estimation of T60 and clarity (C50), two highly correlated parameters, both physically and perceptually. While DRR only measures the direct-to-reverberation ratio, C50 includes early reflections, which also support speech intelligibility. Saini et al.~\cite{Saini2023,saini2024end} proposed a model that estimates both T60 and C50 at sub-band and full-band levels. They use these estimates to select binaural room impulse responses (BRIRs) for spatial reproduction, linking these metrics directly to binaural rendering. Götz et al.~\cite{Gotz2023a} extended this approach to dynamic acoustic environments, with a system capable of real-time adaptation.

More comprehensive models have been proposed for the simultaneous estimation of multiple acoustic parameters. Coldenhoff et al.~\cite{Coldenhoff2024} estimated T60, C50, DRR, and speech quality measures across five microphones, enabling comparisons with single-channel performance. Meng et al.~\cite{Meng2025} used CNNs with first-order Ambisonics (FOA) input to estimate T60, DRR, and C50 in sub-bands, demonstrating the advantages of spatial input formats. Srivastava et al.~\cite{Srivastava2021} estimated parameters such as room volume, surface area, and absorption using multiple two-channel speech recordings. The recent multitask model by Wang et al.~\cite{Wang2025} combines attention and convolutional layers to jointly estimate a wide range of room and geometric parameters, including T60, C50, C80, Definition (D50), early decay time (EDT), center time (Ts), and occupancy, all from single-channel, noisy speech. The results show that the estimation accuracy of several parameters is comparable to or better than the just noticeable difference (JND), highlighting the potential of these estimates for use in spatial perception-based loss functions.  

The final group of studies proposes approaches that are based on embeddings. These models project reverberant signals into latent spaces that encode room acoustics. Bitterman et al.~\cite{Bitterman2024} applied contrastive learning to jointly embed reverberant speech and room impulse responses for room shape classification. Götz et al.~\cite{Gotz2023b} extended this approach to embeddings capable of predicting T60, C50, and volume. Omran et al.~\cite{Omran2023} developed a method to disentangle speech from room acoustics in the embedding space, while Götz et al.~\cite{Gotz2024} introduced task-agnostic embeddings for parameter estimation. Dumpala et al.~\cite{dumpala24_interspeech} proposed embeddings that separate the source from the background and demonstrated their use for estimating C50, T60, DRR, and room volume.

Taken together, the reviewed literature shows significant progress in the blind estimation of room acoustics parameters and the use of embeddings to characterize reverberant environments. Most models rely on single-channel inputs and employ supervised training using simulated or measured data. These contributions offer promising building blocks for perceptually informed loss functions. However, a key gap remains: none of these models have been explicitly integrated into neural network training objectives for spatial audio applications. Their potential for perceptually motivated loss design—particularly in tasks involving reverberant binaural rendering — remains untapped. Bridging this gap could enable training frameworks that directly optimize with respect to room acoustics perception, thereby improving the realism and robustness of spatial audio systems.

\section{Conclusion}

This review examined recent work on loss functions that aim to capture perceptual aspects of spatial audio, with a focus on binaural reproduction. While significant progress has been made in incorporating localization cues and overall quality metrics into optimization objectives, other perceptually important dimensions—such as source distance and room acoustics—remain less developed in loss function design.

This review suggests several possible directions for future research:
\begin{itemize}
  \item Developing perceptually grounded loss functions for quality dimensions such as source distance perception and room acoustics perception.
  \item Transforming existing auditory models and spatial quality measures into differentiable forms usable for neural network training.
  \item Integrating psychoacoustic principles, such as nonlinear frequency resolution and just-noticeable differences, into loss formulations.
  \item Review loss functions and measures related to SAQI qualities not covered in this paper.
\end{itemize}

Overall, advancing loss function design in spatial audio remains a critical step toward creating more accurate, robust, and perceptually convincing learning-based systems.

\bibliographystyle{IEEEtran}

\bibliography{references}

\begin{thebibliography}{10}
\providecommand{\url}[1]{#1}
\csname url@samestyle\endcsname
\providecommand{\newblock}{\relax}
\providecommand{\bibinfo}[2]{#2}
\providecommand{\BIBentrySTDinterwordspacing}{\spaceskip=0pt\relax}
\providecommand{\BIBentryALTinterwordstretchfactor}{4}
\providecommand{\BIBentryALTinterwordspacing}{\spaceskip=\fontdimen2\font plus
\BIBentryALTinterwordstretchfactor\fontdimen3\font minus \fontdimen4\font\relax}
\providecommand{\BIBforeignlanguage}[2]{{%
\expandafter\ifx\csname l@#1\endcsname\relax
\typeout{** WARNING: IEEEtran.bst: No hyphenation pattern has been}%
\typeout{** loaded for the language `#1'. Using the pattern for}%
\typeout{** the default language instead.}%
\else
\language=\csname l@#1\endcsname
\fi
#2}}
\providecommand{\BIBdecl}{\relax}
\BIBdecl

\bibitem{lindau2014SAQI}
\BIBentryALTinterwordspacing
A.~Lindau, V.~Erbes, S.~Lepa, H.-J. Maempel, F.~Brinkman, and S.~Weinzierl, ``A spatial audio quality inventory ({SAQI}),'' \emph{Acta Acustica united with Acustica}, vol. 100, no.~5, pp. 984--994, 2014. [Online]. Available: \url{https://doi.org/10.3813/AAA.918778}
\BIBentrySTDinterwordspacing

\bibitem{brandenburg2019perceptual}
\BIBentryALTinterwordspacing
K.~Brandenburg, B.~Fiedler, G.~Fischer, F.~Klein, A.~Neidhardt, C.~Schneiderwind, U.~Sloma, C.~Stirnat, and S.~Werner, ``Perceptual aspects in spatial audio processing,'' in \emph{Proceedings of the 23rd International Congress on Acoustics: integrating 4th EAA Euroregio 2019: 9-13 September 2019 in Aachen, Germany}, 2019, pp. 3354--3360. [Online]. Available: \url{https://doi.org/10.18154/RWTH-CONV-239329}
\BIBentrySTDinterwordspacing

\bibitem{Hacihabiboglu2017Review}
\BIBentryALTinterwordspacing
H.~Hacihabiboglu, E.~De~Sena, Z.~Cvetkovic, J.~Johnston, and J.~O. Smith~III, ``Perceptual spatial audio recording, simulation, and rendering: An overview of spatial-audio techniques based on psychoacoustics,'' \emph{IEEE Signal Processing Magazine}, vol.~34, no.~3, pp. 36--54, 2017. [Online]. Available: \url{https://doi.org/10.1109/MSP.2017.2666081}
\BIBentrySTDinterwordspacing

\bibitem{Llado2025}
P.~Llad{\'o}, A.~Neidhardt, F.~Brinkmann, and E.~De~Sena, ``Spatial {{Audio Models Inventory}} to {{Cover The Attributes}} from the {{Spatial Audio Quality Inventory}},'' in \emph{Forum {{Acusticum}} ({{Accepted}} for Publication)}, Malaga, Spain, Jun. 2025.

\bibitem{Fantini2025}
\BIBentryALTinterwordspacing
D.~Fantini, M.~Geronazzo, F.~Avanzini, and S.~Ntalampiras, ``A {{Survey}} on {{Machine Learning Techniques}} for {{Head-Related Transfer Function Individualization}},'' \emph{IEEE Open Journal of Signal Processing}, Jan. 2025. [Online]. Available: \url{https://ieeexplore.ieee.org/abstract/document/10836943}
\BIBentrySTDinterwordspacing

\bibitem{cobos2022Review}
\BIBentryALTinterwordspacing
M.~Cobos, J.~Ahrens, K.~Kowalczyk, and A.~Politis, ``An overview of machine learning and other data-based methods for spatial audio capture, processing, and reproduction,'' \emph{EURASIP Journal on Audio, Speech, and Music Processing}, vol.~10, 2022. [Online]. Available: \url{https://doi.org/10.1186/s13636-022-00242-x}
\BIBentrySTDinterwordspacing

\bibitem{BRADLEY2011}
\BIBentryALTinterwordspacing
J.~Bradley, ``Review of objective room acoustics measures and future needs,'' \emph{Applied Acoustics}, vol.~72, no.~10, pp. 713--720, 2011. [Online]. Available: \url{https://www.sciencedirect.com/science/article/pii/S0003682X1100096X}
\BIBentrySTDinterwordspacing

\bibitem{Weinzierl2018RAQI}
\BIBentryALTinterwordspacing
S.~Weinzierl, S.~Lepa, and D.~Ackermann, ``A measuring instrument for the auditory perception of rooms: The room acoustical quality inventory ({RAQI}),'' \emph{The Journal of the Acoustical Society of America}, vol. 144, no.~3, pp. 1245--1257, 09 2018. [Online]. Available: \url{https://doi.org/10.1121/1.5051453}
\BIBentrySTDinterwordspacing

\bibitem{rafaely2022Review}
\BIBentryALTinterwordspacing
{Rafaely, Boaz}, {Tourbabin, Vladimir}, {Habets, Emanuel}, {Ben-Hur, Zamir}, {Lee, Hyunkook}, {Gamper, Hannes}, {Arbel, Lior}, {Birnie, Lachlan}, {Abhayapala, Thushara}, and {Samarasinghe, Prasanga}, ``Spatial audio signal processing for binaural reproduction of recorded acoustic scenes – review and challenges,'' \emph{Acta Acust.}, vol.~6, p.~47, 2022. [Online]. Available: \url{https://doi.org/10.1051/aacus/2022040}
\BIBentrySTDinterwordspacing

\bibitem{Rumsey2005}
F.~Rumsey, S.~Zieli{\'n}ski, R.~Kassier, and S.~Bech, ``On the relative importance of spatial and timbral fidelities in judgments of degraded multichannel audio quality,'' \emph{J. Acoust. Soc. Am.}, vol. 118, no.~2, pp. 968--976, Aug. 2005.

\bibitem{LeBagousse2012}
S.~Le~Bagousse, M.~Paquier, and C.~Colomes, ``Assessment of spatial audio quality based on sound attributes,'' in \emph{Acoustics}, Nantes, France, Apr. 2012, pp. 873--877.

\bibitem{Majdak2022}
\BIBentryALTinterwordspacing
P.~Majdak, C.~Hollomey, and R.~Baumgartner, ``{{AMT}} 1.x: {{A}} toolbox for reproducible research in auditory modeling,'' \emph{Acta Acust.}, vol.~6, p.~19, May 2022. [Online]. Available: \url{https://doi.org/10.1051/aacus/2022011}
\BIBentrySTDinterwordspacing

\bibitem{paszke2019pytorch}
\BIBentryALTinterwordspacing
A.~Paszke, S.~Gross, F.~Massa, A.~Lerer, J.~Bradbury, G.~Chanan, T.~Killeen, Z.~Lin, N.~Gimelshein, L.~Antiga, A.~Desmaison, A.~Kopf, E.~Yang, Z.~DeVito, M.~Raison, A.~Tejani, S.~Chilamkurthy, B.~Steiner, L.~Fang, J.~Bai, and S.~Chintala, ``Pytorch: An imperative style, high-performance deep learning library,'' in \emph{Advances in Neural Information Processing Systems (NeurIPS)}, vol.~32.\hskip 1em plus 0.5em minus 0.4em\relax Curran Associates, Inc., 2019. [Online]. Available: \url{https://doi.org/10.48550/arXiv.1912.01703}
\BIBentrySTDinterwordspacing

\bibitem{pytorch2024}
{PyTorch Developers}, ``Pytorch documentation,'' \url{https://pytorch.org/docs/stable/notes/autograd.html}, 2024, accessed: 2025-06-04.

\bibitem{Jiang2023a}
\BIBentryALTinterwordspacing
G.~Jiang, L.~Villemoes, and A.~Biswas, ``Generative {{Machine Listener}},'' in \emph{155th {{AES Convention}}}, New York, NY, USA, Oct. 2023. [Online]. Available: \url{https://arxiv.org/pdf/2308.09493}
\BIBentrySTDinterwordspacing

\bibitem{manocha2022saqam}
\BIBentryALTinterwordspacing
P.~Manocha, A.~Kumar, B.~Xu, A.~Menon, I.~D. Gebru, V.~K. Ithapu, and P.~Calamia, ``{SAQAM}: Spatial audio quality assessment metric,'' in \emph{Proc. Interspeech 2022}, 2022, pp. 649--653. [Online]. Available: \url{https://doi.org/10.21437/Interspeech.2022-406}
\BIBentrySTDinterwordspacing

\bibitem{Zheng2025}
\BIBentryALTinterwordspacing
Y.~Zheng, J.~Yao, X.~Deng, Y.~Yang, R.~Liao, W.~Tu, and C.~Lin, ``{{HAPG-SAQAM}}: {{Human Auditory Perception Guided Spatial Audio Quality Assessment Metric}},'' in \emph{{{IEEE Int}}. {{Conf}}. {{Acoustics}}, {{Speech}} and {{Signal Processing}} ({{ICASSP}})}, Apr. 2025. [Online]. Available: \url{https://ieeexplore.ieee.org/document/10889173}
\BIBentrySTDinterwordspacing

\bibitem{eurich2024computationally}
\BIBentryALTinterwordspacing
B.~Eurich, S.~D. Ewert, M.~Dietz, and T.~Biburger, ``A computationally efficient model for combined assessment of monaural and binaural audio quality,'' \emph{J. Audio Eng. Soc}, vol.~72, no.~9, pp. 536--551, 2024. [Online]. Available: \url{https://doi.org/10.17743/jaes.2022.0161}
\BIBentrySTDinterwordspacing

\bibitem{Schafer2013}
\BIBentryALTinterwordspacing
M.~Sch{\"a}fer, M.~Bahram, and P.~Vary, ``An extension of the {{PEAQ}} measure by a binaural hearing model,'' in \emph{{{IEEE Int}}. {{Conf}}. {{Acoustics}}, {{Speech}} and {{Signal Processing}} ({{ICASSP}})}, Vancouver, BC, Canada, May 2013, pp. 8164--8168. [Online]. Available: \url{https://ieeexplore.ieee.org/abstract/document/6639256}
\BIBentrySTDinterwordspacing

\bibitem{Delgado2023}
P.~Delgado and J.~Herre, ``Design choices in a binaural perceptual model for improved objective spatial audio quality assessment,'' in \emph{155th {{AES Convention}}}, New York, NY, USA, Oct. 2023.

\bibitem{flessner2017assessment}
\BIBentryALTinterwordspacing
J.-H. Fle{\ss}ner, R.~Huber, and S.~D. Ewert, ``Assessment and prediction of binaural aspects of audio quality,'' \emph{Journal of the Audio Engineering Society}, vol.~65, no.~11, pp. 929--942, 2017. [Online]. Available: \url{https://doi.org/10.17743/jaes.2017.0037}
\BIBentrySTDinterwordspacing

\bibitem{Flessner2019}
\BIBentryALTinterwordspacing
J.-H. Fle{\ss}ner, T.~Biberger, and S.~D. Ewert, ``Subjective and {{Objective Assessment}} of {{Monaural}} and {{Binaural Aspects}} of {{Audio Quality}},'' \emph{IEEE/ACM Transactions on Audio, Speech, and Language Processing}, vol.~27, no.~7, pp. 1112--1125, Jul. 2019. [Online]. Available: \url{https://ieeexplore.ieee.org/abstract/document/8708700}
\BIBentrySTDinterwordspacing

\bibitem{Narbutt2018}
\BIBentryALTinterwordspacing
M.~Narbutt, A.~Allen, J.~Skoglund, M.~Chinen, and A.~Hines, ``Ambiqual - a full reference objective quality metric for ambisonic spatial audio,'' in \emph{2018 Tenth International Conference on Quality of Multimedia Experience (QoMEX)}, 2018, pp. 1--6. [Online]. Available: \url{https://doi.org/10.1109/QoMEX.2018.8463408}
\BIBentrySTDinterwordspacing

\bibitem{Narbutt2020}
\BIBentryALTinterwordspacing
M.~Narbutt, J.~Skoglund, A.~Allen, M.~Chinen, D.~Barry, and A.~Hines, ``Ambiqual: Towards a quality metric for headphone rendered compressed ambisonic spatial audio,'' \emph{Applied Sciences}, vol.~10, no.~9, 2020. [Online]. Available: \url{https://doi.org/10.3390/app10093188}
\BIBentrySTDinterwordspacing

\bibitem{Hogg2025}
A.~O.~T. Hogg, R.~Barumerli, R.~Daugintis, K.~C. Poole, F.~Brinkmann, L.~Picinali, and M.~Geronazzo, ``Listener {{Acoustic Personalisation Challenge}} - {{LAP24}}: {{Head-Related Transfer Function Upsampling}},'' \emph{IEEE Open Journal of Signal Processing}, 2025 (Submitted).

\bibitem{Yao2024a}
\BIBentryALTinterwordspacing
D.~Yao, J.~Zhao, Y.~Liang, Y.~Wang, J.~Gu, M.~Jia, H.~Lee, and J.~Li, ``Perceptually enhanced spectral distance metric for head-related transfer function quality prediction,'' \emph{JASA}, vol. 156, no.~6, pp. 4133--4152, Dec. 2024. [Online]. Available: \url{https://doi.org/10.1121/10.0034632}
\BIBentrySTDinterwordspacing

\bibitem{McKenzie2025}
\BIBentryALTinterwordspacing
T.~McKenzie and F.~Brinkmann, ``Toward an {{Improved Auditory Model}} for {{Predicting Binaural Coloration}},'' \emph{J. Audio Eng. Soc.}, vol.~73, no.~3, pp. 115--126, Mar. 2025. [Online]. Available: \url{https://aes2.org/publications/elibrary-page/?id=22807}
\BIBentrySTDinterwordspacing

\bibitem{Krause2021}
\BIBentryALTinterwordspacing
D.~A. Krause, A.~Politis, and A.~Mesaros, ``Joint direction and proximity classification of overlapping sound events from binaural audio,'' in \emph{2021 IEEE Workshop on Applications of Signal Processing to Audio and Acoustics (WASPAA)}, 2021, pp. 331--335. [Online]. Available: \url{https://doi.org/10.1109/WASPAA52581.2021.9632775}
\BIBentrySTDinterwordspacing

\bibitem{panah2025binaqual}
\BIBentryALTinterwordspacing
D.~S. Panah, D.~Barry, A.~Ragano, J.~Skoglund, and A.~Hines, ``Binaqual: A full-reference objective localization similarity metric for binaural audio,'' \emph{arXiv preprint arXiv:2505.11915}, 2025. [Online]. Available: \url{https://arxiv.org/abs/2505.11915}
\BIBentrySTDinterwordspacing

\bibitem{manocha2021dplm}
\BIBentryALTinterwordspacing
P.~Manocha, A.~Kumar, B.~Xu, A.~Menon, I.~D. Gebru, V.~K. Ithapu, and P.~Calamia, ``Dplm: A deep perceptual spatial-audio localization metric,'' in \emph{2021 IEEE Workshop on Applications of Signal Processing to Audio and Acoustics (WASPAA)}.\hskip 1em plus 0.5em minus 0.4em\relax IEEE, 2021, pp. 6--10. [Online]. Available: \url{https://ieeexplore.ieee.org/abstract/document/9632781}
\BIBentrySTDinterwordspacing

\bibitem{francl2022deep}
A.~Francl and J.~H. McDermott, ``Deep neural network models of sound localization reveal how perception is adapted to real-world environments,'' \emph{Nature human behaviour}, vol.~6, no.~1, pp. 111--133, 2022.

\bibitem{berebi2023imagls}
\BIBentryALTinterwordspacing
O.~Berebi, Z.~Ben-Hur, D.~L. Alon, and B.~Rafaely, ``{iMagLS}: Interaural level difference with magnitude least-squares loss for optimized first-order head-related transfer function,'' in \emph{10th Convention of the European Acoustics Association, EAA 2023}.\hskip 1em plus 0.5em minus 0.4em\relax European Acoustics Association, EAA, 2023. [Online]. Available: \url{https://dael.euracoustics.org/confs/fa2023/data/articles/000678.pdf}
\BIBentrySTDinterwordspacing

\bibitem{berebi2024feasibility}
\BIBentryALTinterwordspacing
------, ``Feasibility of {iMagLS-BSM-ILD} informed binaural signal matching with arbitrary microphone arrays,'' in \emph{2024 18th International Workshop on Acoustic Signal Enhancement (IWAENC)}.\hskip 1em plus 0.5em minus 0.4em\relax IEEE, 2024, pp. 429--433. [Online]. Available: \url{https://ieeexplore.ieee.org/abstract/document/10694218}
\BIBentrySTDinterwordspacing

\bibitem{berebi2025bsm}
\BIBentryALTinterwordspacing
------, ``{BSM-iMagLS}: {ILD} informed binaural signal matching for reproduction with head-mounted microphone arrays,'' \emph{arXiv preprint arXiv:2501.18227}, 2025. [Online]. Available: \url{https://doi.org/10.48550/arXiv.2501.18227}
\BIBentrySTDinterwordspacing

\bibitem{10889034}
\BIBentryALTinterwordspacing
O.~Berebi, F.~Brinkmann, S.~Weinzierl, and B.~Rafaely, ``Ambisonics binaural rendering via masked magnitude least squares,'' in \emph{ICASSP 2025 - 2025 IEEE International Conference on Acoustics, Speech and Signal Processing (ICASSP)}, 2025, pp. 1--5. [Online]. Available: \url{https://ieeexplore.ieee.org/document/10889034}
\BIBentrySTDinterwordspacing

\bibitem{tokala2024binaural}
\BIBentryALTinterwordspacing
V.~Tokala, E.~Grinstein, M.~Brookes, S.~Doclo, J.~Jensen, and P.~A. Naylor, ``Binaural speech enhancement using deep complex convolutional transformer networks,'' in \emph{ICASSP 2024-2024 IEEE International Conference on Acoustics, Speech and Signal Processing (ICASSP)}.\hskip 1em plus 0.5em minus 0.4em\relax IEEE, 2024, pp. 681--685. [Online]. Available: \url{https://ieeexplore.ieee.org/abstract/document/10447090}
\BIBentrySTDinterwordspacing

\bibitem{neudekinvestigation}
\BIBentryALTinterwordspacing
D.~Neudek, B.~Stodt, S.~Getzmann, and R.~Martin, ``Investigation of binaural distance estimation with artificial neural networks trained on simulated data models.'' [Online]. Available: \url{https://pub.dega-akustik.de/DAS-DAGA_2025/files/upload/paper/355.pdf}
\BIBentrySTDinterwordspacing

\bibitem{gburrek2024diminishing}
\BIBentryALTinterwordspacing
T.~Gburrek, A.~Meise, J.~Schmalenstroeer, and R.~Haeb-Umbach, ``Diminishing domain mismatch for dnn-based acoustic distance estimation via stochastic room reverberation models,'' in \emph{2024 18th International Workshop on Acoustic Signal Enhancement (IWAENC)}.\hskip 1em plus 0.5em minus 0.4em\relax IEEE, 2024, pp. 279--283. [Online]. Available: \url{https://ieeexplore.ieee.org/abstract/document/10694103}
\BIBentrySTDinterwordspacing

\bibitem{neri2024speaker}
\BIBentryALTinterwordspacing
M.~Neri, A.~Politis, D.~A. Krause, M.~Carli, and T.~Virtanen, ``Speaker distance estimation in enclosures from single-channel audio,'' \emph{IEEE/ACM Transactions on Audio, Speech, and Language Processing}, vol.~32, pp. 2242--2254, 2024. [Online]. Available: \url{https://ieeexplore.ieee.org/abstract/document/10480456}
\BIBentrySTDinterwordspacing

\bibitem{ITU-R1534-3}
{ITU-R BS.1534-3}, \emph{Methods for the Subjective Assessment of Intermediate Quality Level of Audio Systems}.\hskip 1em plus 0.5em minus 0.4em\relax Geneva, Switzerland: ITU, 2015.

\bibitem{zotter2019ambisonics}
\BIBentryALTinterwordspacing
F.~Zotter and M.~Frank, \emph{Ambisonics: A Practical 3D Audio Theory for Recording, Studio Production, Sound Reinforcement, and Virtual Reality}, 1st~ed., ser. Springer Topics in Signal Processing.\hskip 1em plus 0.5em minus 0.4em\relax Cham, Switzerland: Springer International Publishing, 2019, vol.~19. [Online]. Available: \url{https://doi.org/10.1007/978-3-030-17207-7}
\BIBentrySTDinterwordspacing

\bibitem{cox2001extracting}
T.~J. Cox, F.~Li, and P.~Darlington, ``Extracting room reverberation time from speech using artificial neural networks,'' \emph{Journal of the Audio Engineering Society}, vol.~49, no.~4, pp. 219--230, 2001.

\bibitem{Gamper2018}
\BIBentryALTinterwordspacing
H.~Gamper and I.~J. Tashev, ``Blind reverberation time estimation using a convolutional neural network,'' in \emph{2018 16th International Workshop on Acoustic Signal Enhancement (IWAENC)}, 2018, pp. 136--140. [Online]. Available: \url{https://doi.org/10.1109/IWAENC.2018.8521241}
\BIBentrySTDinterwordspacing

\bibitem{Li2021}
\BIBentryALTinterwordspacing
Y.~Li, Y.~Liu, and D.~S. Williamson, ``On loss functions for deep-learning based {T60} estimation,'' in \emph{ICASSP 2021 - 2021 IEEE International Conference on Acoustics, Speech and Signal Processing (ICASSP)}, 2021, pp. 486--490. [Online]. Available: \url{https://doi.org/10.1109/ICASSP39728.2021.9414442}
\BIBentrySTDinterwordspacing

\bibitem{Li2023}
\BIBentryALTinterwordspacing
------, ``A composite {T60} regression and classification approach for speech dereverberation,'' \emph{IEEE/ACM Transactions on Audio, Speech, and Language Processing}, vol.~31, pp. 1013--1023, 2023. [Online]. Available: \url{https://doi.org/10.1109/TASLP.2023.3245423}
\BIBentrySTDinterwordspacing

\bibitem{Deng2020}
\BIBentryALTinterwordspacing
S.~Deng, W.~Mack, and E.~A.~P. Habets, ``Online blind reverberation time estimation using {CRNNs},'' in \emph{Proc. Interspeech}, 2020, pp. 5061--5065. [Online]. Available: \url{https://doi.org/10.21437/Interspeech.2020-2156}
\BIBentrySTDinterwordspacing

\bibitem{Genovese2019}
\BIBentryALTinterwordspacing
A.~F. Genovese, H.~Gamper, V.~Pulkki, N.~Raghuvanshi, and I.~J. Tashev, ``Blind room volume estimation from single-channel noisy speech,'' in \emph{Proc. IEEE Intl. Conf. on Acoustics, Speech and Signal Processing (ICASSP)}, 2019, pp. 231--235. [Online]. Available: \url{https://doi.org/10.1109/ICASSP.2019.8682951}
\BIBentrySTDinterwordspacing

\bibitem{Mack2020}
\BIBentryALTinterwordspacing
W.~Mack, S.~Deng, and E.~A.~P. Habets, ``Single-channel blind direct-to-reverberation ratio estimation using masking,'' in \emph{Proc. Interspeech}, 2020, pp. 5066--5070. [Online]. Available: \url{http://dx.doi.org/10.21437/Interspeech.2020-2171}
\BIBentrySTDinterwordspacing

\bibitem{Eaton2016}
\BIBentryALTinterwordspacing
J.~Eaton, N.~D. Gaubitch, A.~H. Moore, and P.~A. Naylor, ``Estimation of room acoustic parameters: The {ACE} challenge,'' \emph{IEEE/ACM Transactions on Audio, Speech, and Language Processing}, vol.~24, no.~10, pp. 1681--1693, 2016. [Online]. Available: \url{https://doi.org/10.1109/TASLP.2016.2577502}
\BIBentrySTDinterwordspacing

\bibitem{Looney2020}
\BIBentryALTinterwordspacing
D.~Looney and N.~D. Gaubitch, ``Joint estimation of acoustic parameters from single-microphone speech observations,'' in \emph{ICASSP 2020 - 2020 IEEE International Conference on Acoustics, Speech and Signal Processing (ICASSP)}, 2020, pp. 431--435. [Online]. Available: \url{https://doi.org/10.1109/ICASSP40776.2020.9054532}
\BIBentrySTDinterwordspacing

\bibitem{Saini2023}
\BIBentryALTinterwordspacing
S.~Saini and J.~Peissig, ``Blind room acoustic parameters estimation using mobile audio transformer,'' in \emph{2023 IEEE Workshop on Applications of Signal Processing to Audio and Acoustics (WASPAA)}, 2023, pp. 1--5. [Online]. Available: \url{https://doi.org/10.1109/WASPAA58266.2023.10248186}
\BIBentrySTDinterwordspacing

\bibitem{saini2024end}
\BIBentryALTinterwordspacing
S.~Saini, I.~Engel, and J.~Peissig, ``An end-to-end approach for blindly rendering a virtual sound source in an audio augmented reality environment,'' \emph{EURASIP Journal on Audio, Speech, and Music Processing}, vol. 2024, no.~1, p.~16, 2024. [Online]. Available: \url{https://asmp-eurasipjournals.springeropen.com/articles/10.1186/s13636-024-00338-6}
\BIBentrySTDinterwordspacing

\bibitem{Gotz2023a}
\BIBentryALTinterwordspacing
P.~Götz, C.~Tuna, A.~Walther, and E.~A.~P. Habets, ``Online reverberation time and clarity estimation in dynamic acoustic conditions,'' \emph{The Journal of the Acoustical Society of America}, vol. 153, no.~6, pp. 3532--3542, 2023. [Online]. Available: \url{https://doi.org/10.1121/10.0019804}
\BIBentrySTDinterwordspacing

\bibitem{Coldenhoff2024}
\BIBentryALTinterwordspacing
J.~Coldenhoff, A.~Harper, P.~Kendrick, T.~Stojkovic, and M.~Cernak, ``Multi-channel {MORSA}: Mean opinion score and room acoustics estimation using simulated data and a teacher model,'' in \emph{ICASSP 2024 - 2024 IEEE International Conference on Acoustics, Speech and Signal Processing (ICASSP)}, 2024, pp. 381--385. [Online]. Available: \url{https://doi.org/10.1109/ICASSP48485.2024.10447841}
\BIBentrySTDinterwordspacing

\bibitem{Meng2025}
\BIBentryALTinterwordspacing
H.~Meng, J.~Breebaar, J.~Stoddard, V.~Sethu, and E.~Ambikairajah, ``Blind estimation of sub-band acoustic parameters from ambisonics recordings using spectro-spatial covariance features,'' in \emph{ICASSP 2025 - 2025 IEEE International Conference on Acoustics, Speech and Signal Processing (ICASSP)}, 2025, pp. 1--5. [Online]. Available: \url{https://doi.org/10.1109/ICASSP49660.2025.10887842}
\BIBentrySTDinterwordspacing

\bibitem{Srivastava2021}
\BIBentryALTinterwordspacing
P.~Srivastava, A.~Deleforge, and E.~Vincent, ``Blind room parameter estimation using multiple-multichannel speech recordings,'' in \emph{IEEE Workshop on Applications of Signal Processing to Audio and Acoustics (WASPAA)}, 2021, pp. 226--230. [Online]. Available: \url{https://doi.org/10.1109/WASPAA52581.2021.9632778}
\BIBentrySTDinterwordspacing

\bibitem{Wang2025}
\BIBentryALTinterwordspacing
L.~Wang, Y.~Lu, Z.~Gao, K.~Li, J.~Huang, Y.~Kong, and S.~Okada, ``Berp: A blind estimator of room parameters for single-channel noisy speech signals,'' \emph{IEEE Transactions on Audio, Speech and Language Processing}, 2025, early Access. [Online]. Available: \url{https://doi.org/10.1109/TASLPRO.2025.3574849}
\BIBentrySTDinterwordspacing

\bibitem{Bitterman2024}
\BIBentryALTinterwordspacing
J.~Bitterman, D.~Levi, H.~H. Diamandi, S.~Gannot, and T.~Rosenwein, ``{RevRIR}: Joint reverberant speech and room impulse response embedding using contrastive learning with application to room shape classification,'' in \emph{Proc. Interspeech}, 2024, pp. 3280--3284. [Online]. Available: \url{https://doi.org/10.21437/Interspeech.2024-1951}
\BIBentrySTDinterwordspacing

\bibitem{Gotz2023b}
\BIBentryALTinterwordspacing
P.~Götz, C.~Tuna, A.~Walther, and E.~A.~P. Habets, ``Contrastive representation learning for acoustic parameter estimation,'' in \emph{Proc. IEEE Intl. Conf. on Acoustics, Speech and Signal Processing (ICASSP)}, 2023, pp. 1--5. [Online]. Available: \url{https://doi.org/10.1109/ICASSP49357.2023.10095279}
\BIBentrySTDinterwordspacing

\bibitem{Omran2023}
\BIBentryALTinterwordspacing
A.~Omran, N.~Zeghidour, Z.~Borsos, F.~de~Chaumont~Quitry, M.~Slaney, and M.~Tagliasacchi, ``Disentangling speech from surroundings with neural embeddings,'' in \emph{Proc. IEEE Intl. Conf. on Acoustics, Speech and Signal Processing (ICASSP)}, 2023, pp. 1--5. [Online]. Available: \url{https://doi.org/10.1109/ICASSP49357.2023.10096435}
\BIBentrySTDinterwordspacing

\bibitem{Gotz2024}
\BIBentryALTinterwordspacing
P.~Götz, C.~Tuna, A.~Brendel, A.~Walther, and E.~A.~P. Habets, ``Blind acoustic parameter estimation through task-agnostic embeddings using latent approximations,'' in \emph{Proc. Intl. Workshop on Acoustic Signal Enhancement (IWAENC)}, 2024, pp. 289--293. [Online]. Available: \url{https://doi.org/10.1109/IWAENC61483.2024.10694126}
\BIBentrySTDinterwordspacing

\bibitem{dumpala24_interspeech}
\BIBentryALTinterwordspacing
S.~H. Dumpala, D.~Sharma, C.~{Shama Sastry}, S.~Kruchinin, J.~Fosburgh, and P.~A. Naylor, ``{XANE}: e{X}plainable {A}coustic {N}eural {E}mbeddings,'' in \emph{Interspeech 2024}, 2024, pp. 3824--3828. [Online]. Available: \url{https://doi.org/10.21437/Interspeech.2024-1229}
\BIBentrySTDinterwordspacing

\end{thebibliography}

\end{document}